\documentclass[review,times, 3p]{elsarticle}
\usepackage{amsmath}
\usepackage{graphicx}
\usepackage{float}
\usepackage{xcolor}
\usepackage{graphicx}% Include figure files
\usepackage{dcolumn}% Align table columns on decimal point
\usepackage{bm}% bold math
\usepackage{subcaption}
\usepackage{caption}
\usepackage{cuted}
\usepackage{placeins} % used to allow \floatbarrier
\usepackage{lineno}
\usepackage{soul}
\usepackage[ruled,vlined]{algorithm2e}

\usepackage[T1]{fontenc}

\usepackage{hyperref}% add hypertext capabilities
\hypersetup{colorlinks=true, citecolor=blue, urlcolor=blue, linkcolor=blue}

\def \etal {{\it et al.}}

\newfloat{algo}{th}{alg}
\restylefloat{figure}% ...figure environment contents.

\makeatletter
\newcommand{\removelatexerror}{\let\@latex@error\@gobble}
\makeatother

\begin{document}

\title{Machine learning assisted non-destructive beam profile monitoring}
\author[1,2]{Zhanibek Omarov}
\author[2]{Selcuk Hac{\i}{\" o}mero{\u g}lu\corref{cor1}}
\address[1]{Korea Institute of Science and Technology, Daejeon, 34141, Republic of Korea}
\address[2]{Institute for Basic Science, Daejeon, 34051, Republic of Korea}
\cortext[cor1]{Corresponding author: selcuk.haciomeroglu@gmail.com}

\begin{abstract}
We present a non-destructive beam profile monitoring concept that utilizes
numerical optimization tools, namely genetic algorithm with a gradient
descent-like minimization. The signal picked up by a button BPM includes
information about the transverse profile content of the beam. A genetic
algorithm is used to transform an arbitrary Gaussian beam in such a way that it
eventually reconstructs the transverse position and the shape of the original
beam to match the signal on the BPM electrodes. A case study for the developed
algorithm is proton EDM experiment where conventional beam profile measurements
are not possible. This method allows visualization of fairly distorted beams
with non-Gaussian distributions as well.

\end{abstract}

\begin{keyword}
Non-destructive beam profile monitor \sep Iterative reconstruction \sep Genetic algorithm.
\end{keyword}

\maketitle

%\twocolumn
%\linenumbers
\section{Introduction}

A beam profile monitor is needed for the proton EDM experiment, which aims to
search for the electric dipole moment (EDM) of proton with $10^{-29}~e\cdot$cm
sensitivity \cite{ref:edm_rsi, ref:symm_hyb_ring}. It is based on measurement of
out-of-plane spin precession rate inside a storage ring. This spin precession is
a result of coupling between the main (radial) electric field and spin of
longitudinally polarized beams. The beam with  $1.17\times 10^{8}$ protons per
bunch roughly corresponds to 1.7 mA. For spin coherence related
restrictions, the momentum will be 0.7 G$e$V/$c$ with a spread within $\vert
\Delta p/p \vert =10^{-4}$. Since the coupling between the magnetic dipole
moment and the electric and magnetic fields is orders of magnitude larger than
the EDM coupling, the experiment also requires a strict control of the external
fields. For example, the magnetic fields must be kept below nanoTesla level.
Similarly, the net electric focusing index along the ring must be at the order
of $m=10^{-4}$. A comprehensive investigation of these systematic errors and
their solutions are provided at Ref. \cite{ref:symm_hyb_ring}. In addition to
the field-related restrictions, interaction of the beam with material is also
avoided as it can result in quick depolarization, as well as energy loss that
has an indirect effect on the spin dynamics. The vacuum also must be at the
order of $10^{-10}$ Torr for a good beam lifetime. 

Transverse profile of a particle beam in an accelerator can be obtained in a
number of ways. Most commonly used methods include wire scanners
\cite{ref:wirescanner, ref:flyingwire, ref:rot_wire, ref:swissfel}, laser wires
\cite{ref:laserwire_0, ref:laserwire}, gas ionization methods \cite{ref:ionization,ref:ionization2,ref:res-gas,
ref:res-gas-2}, silicon detectors in combination with gas ionization setups or
scintillating screens \cite{ref:silicondetector, ref:otr_1, ref:otr_2},
transition radiation detectors \cite{ref:transition,ref:transition2}, and so
on. Some of them extract a portion of the beam for sampling, while others rely
on secondary effects like ionization of a residual or injected gas on the beam
path.  

\begin{figure}[tb]
	\centering
	\includegraphics[width=.4\linewidth]{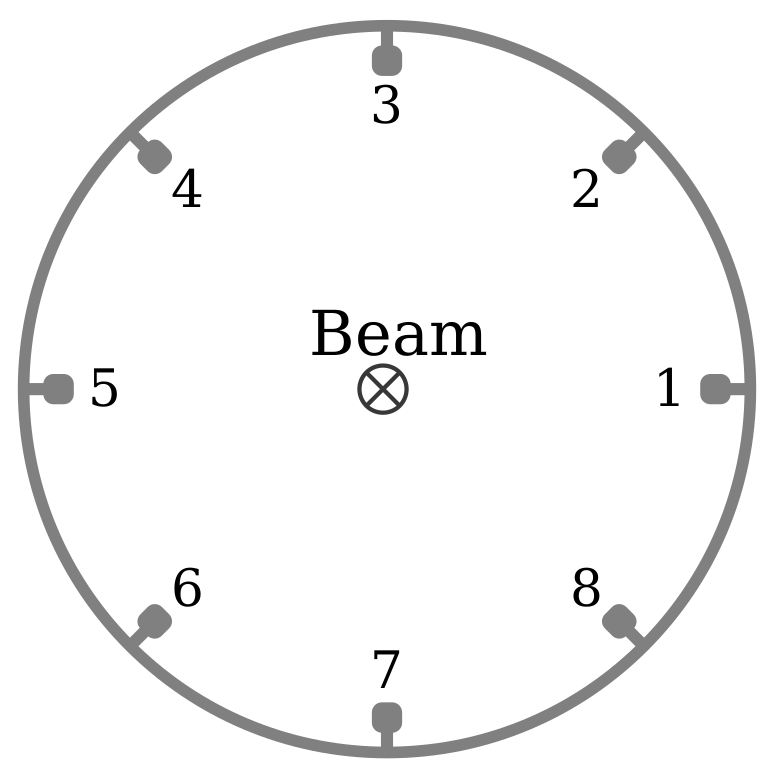}
	\caption{Sketch of a button BPM cross section. The numbers indicate the
	pick-up electrodes.}\label{fig:crossandside}
\end{figure}

Considering all the restrictions mentioned above, wire scanners, gas ionization
methods, and scintillating screens turn out to be destructive methods for this
particular experiment. In this work, we present an alternative transverse beam
profile imaging method, which is based on measurement of the induced magnetic or
electric fields  around the beam. This method requires a hardware that is
similar to a conventional button beam position monitors, only with more number
of probes (Figure \ref{fig:crossandside}). 

%The negligible interaction between the beam and the probes ensures a non-destructive measurement for the proton EDM experiment.

Usage of pick-up electrodes for beam emittance measurement was first proposed by Miller
\etal~in 1983 \cite{ref:miller}. They measured the quadrupole moment of the
transverse beam distribution by using six stripline beam position monitors at
different locations of a linac, and then fitted it with the transport matrix.
Based on the analysis of Miller \etal, it was shown that a stripline-type beam
position monitor can be used for measuring the size of a perpendicular
ellipsoidal beam \cite{ref:suwada1}, and the energy spread \cite{ref:suwada2} as
well. A recent study showed that the beam size measurement sensitivity with this
method can be further improved by using movable pick-up electrodes
\cite{ref:move_pickup}.  In all those studies, the beam was assumed to have a
well-defined shape: Gaussian or elliptical. 

While the transverse beam profile information is embedded in the measured signal
at the electrodes, there are two difficulties with these measurements. As will
be shown below, the multipole moments that can be extracted from the signal on
the electrodes are not sufficient to infer the beam profile without beam size
knowledge. Moreover, it is difficult to estimate the beam profile analytically
if the beam is deformed, i.e., lacks a simple symmetry. We will demonstrate
through simulations that by using appropriate computational tools, it is
possible to visually reconstruct the transverse profile of a distorted beam with
a button BPM. We will not get into the details of the BPM design as it is beyond
the scope of this work.   

This method has similarities with the well-known iterative reconstruction of
computed tomography (CT) images. In a typical application, an approximate sample
structure is obtained analytically from the detector measurements. Then, the
solution is modified iteratively on computer for de-noising and further tuning.
Short introductions to that technique are given at Refs
\cite{ref:iter_reconstr_1, ref:iter_reconstr_2}. One can also refer to Ref
\cite{ref:iter_reconstr_3} and the references therein for more details. Our
method differs by not requiring an analytical estimation for the reconstruction.
Moreover, the iterative modifications are made through ``physically meaningful''
operators for faster convergence. 

%It is worth mentioning that usage of this algorithm is not limited to beam
%profile monitoring. It can as well be utilized at other fields as long as the
%image in question can be obtained from an arbitrary one through discrete
%transformations.

Section \ref{sect:multipole_analysis} presents analytical description of the
problem. Section \ref{sect:method} explains the algorithm for the beam profile
reconstruction, as well as the choice of relevant parameters.  Section
\ref{sect:results} presents the reconstruction performance of this method for
randomly generated beam profiles. Effect of the beam size error is investigated
at \ref{sec:beamsizeappendix}. We state remarks regarding using neural networks for this task at Appendix \ref{sec:neuralnetwork}.

\section{Multipole analysis}{\label{sect:multipole_analysis}}
The beam profile monitor (BPM) in this work includes eight pick-up electrodes,
uniformly distributed around the beam (Figure \ref{fig:crossandside}).  An
incoming beam generates a signal on the electrodes, which depends on its
transverse distribution and the current it carries. The induced current on an
electrode by a delta function line current is given in the cylindrical coordinates
as \cite{ref:miller}
\begin{equation}
\label{eq:current}
J_{\textnormal{image}} (r, \phi, a, \theta) = \frac{I(r,\phi)}{2\pi a} \frac{a^{2} - r^{2}}{a^{2}+r^{2}-2ar\cos(\theta-\phi)},
\end{equation}
where $I$ is the beam current, $a$ and $\phi$ are the radial position and the
azimuthal angle of the electrode, $r$ and $\theta$ are those of the beam,
respectively. Expanding it in powers of $r/a$, one obtains
\begin{equation}
\label{eq:current_gauss_raw}
J_{\textnormal{image}} (r, \phi, a, \theta) = \frac{I(r,\phi)}{2\pi a} \left[ 1+2\sum_{k=1}^{\infty}\left (\frac{r}{a} \right)^{k} \cos k(\theta-\phi) \right].
\end{equation}

This formalism can naturally be extended to an arbitrary beam profile
distribution by integration. For a Gaussian charge distribution
with standard deviations $\sigma_{x}$ and $\sigma_{y}$, centered at $\bar{x}$
and $\bar{y}$, it becomes
\begin{equation}
\label{eq:current_gauss}
J_{\textnormal{image}} (a, \theta) = \frac{I_{\textnormal{beam}}}{4\pi^{2}a\sigma_{x}\sigma_{y}} \iint_{\textnormal{beam}} \left[1 + 2 \sum_{k=1}^{\infty}\left(\frac{r}{a}\right )^{k} \cos k\left(\theta - \phi\right)\right] \\
  \exp \left[\frac{{\left(x-\bar{x}\right)}^{2}}{2\sigma_{x}^2}\right]\exp \left[\frac{{\left(y-\bar{y}\right)}^{2}}{2\sigma_{y}^2}\right] dA,
\end{equation}
where $k$ is the order of the multipoles. Further expanding with $k$ for $r<a$ gives
\begin{equation}
	\begin{split}
	J_{\textnormal{image}} (a, \theta) \approx   \frac{I_{\textnormal{beam}}}{2\pi a} &\Bigg( 1+2 \left[\frac{\bar{x}}{a}\cos\theta + \frac{\bar{y}}{a}\sin\theta \right] 
	 +2\left[ \left(\frac{\sigma_{x}^{2} - \sigma_{y}^{2}}{a^{2}} + \frac{\bar{x}^{2} - \bar{y}^{2}}{a^{2}}\right)\cos 2\theta + 2 \frac{\bar{x}\bar{y}}{a^{2}}\sin 2\theta \right] \\
	& + 2\left[ 3\left(\frac{\sigma_{x}^{2} - \sigma_{y}^{2}}{a^{2}}\right) + \frac{\bar{x}^{2} - \bar{y}^{2}}{a^{2}}\right] \left(\frac{\bar{x}}{a}\cos 3\theta + \frac{\bar{y}}{a}\sin 3\theta\right) + \cdots \Bigg),
	\end{split}
	\label{eq:quadrupole}
\end{equation}
where higher order terms are omitted as an approximation for a pencil beam in which $\sigma_{x,y}$ is small compared to the beam pipe radius~\cite{ref:miller}. Nevertheless, the main operational equation of this work is Equation \ref{eq:current}; hence free of approximations.

It is clear that two beams with different profiles can induce the same signal if
their $\sigma^2=\sigma_x^2-\sigma_y^2$ and $\bar{x}^2-\bar{y}^2$ are the same.
This causes degenerate solutions, which result in arbitrary changes in the size
of the reconstructed beam. This systematic error can be reduced with knowledge
of the beam size, or alternatively  beam emittances ($\epsilon_{x,y}$) and beta
functions ($\beta_{x,y}$). In the proton EDM experiment, the beam size is
expected to be known precisely thanks to the continuous beam extraction. Effect
of the $\epsilon_{x,y}$ and $\beta_{x,y}$ drifts is analyzed in the
\ref{sec:beamsizeappendix}.

\section{Beam profile reconstruction routine}{\label{sect:method}}

\begin{figure}
\centering
\includegraphics[width=0.6\linewidth]{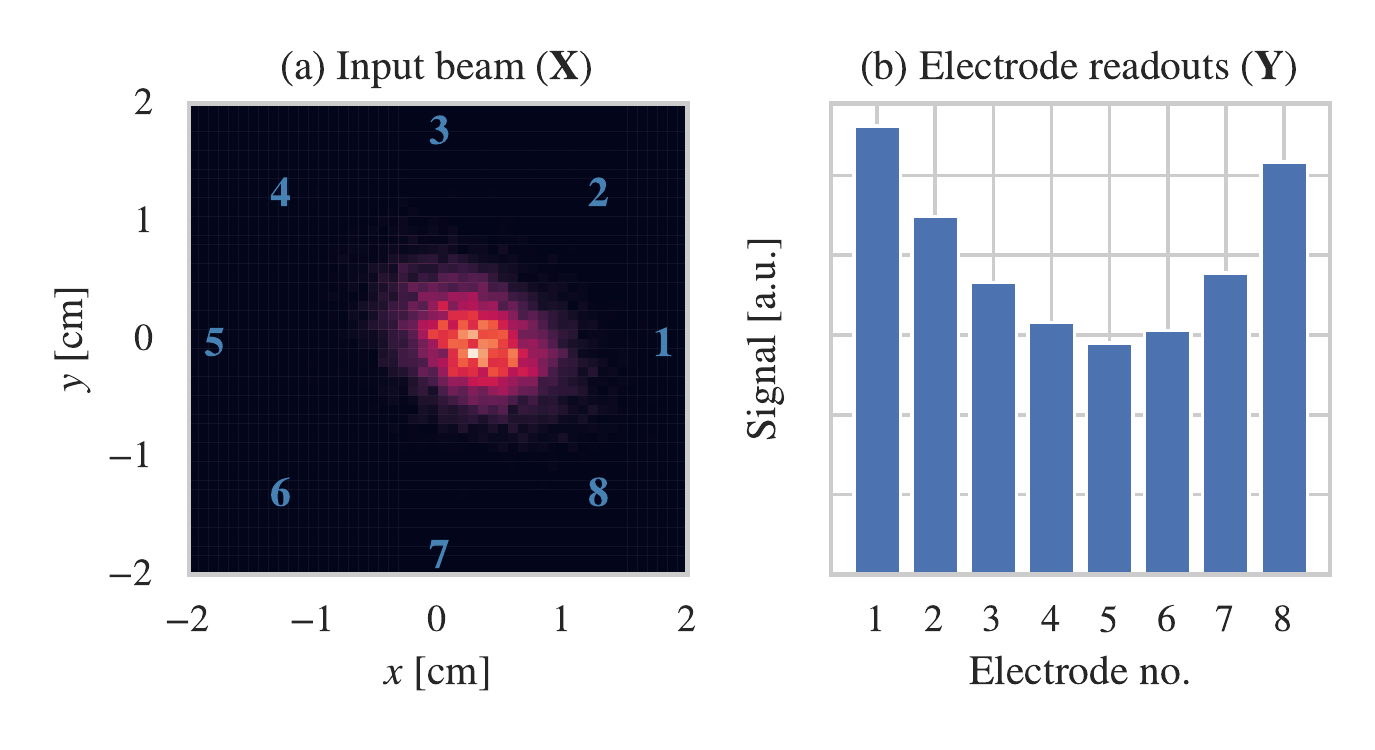}
\caption{(a) A beam profile is represented by a $2 \times N$ matrix \textbf{X}.
The numbers from 1 to 8 represent the locations of the electrodes. The color
code indicates the intensity of particles, where bright colors indicate higher
intensity. (b) The induced signal on the electrodes is represented by an
8-dimensional vector \textbf{Y}.}
\label{fig:transform}
\end{figure}

An arbitrary beam profile with $N$ particles of $(x,y)$ coordinates can be
represented by a $2\times N$ matrix \textbf{X}. The beam induces a signal, which
is represented by an 8-dimensional vector \textbf{Y}, on the eight electrodes of
the BPM (Figure \ref{fig:crossandside}). A sample beam of $N=10^4$ particles and
the corresponding signal on the electrodes are shown in Figure
\ref{fig:transform}.

Before presenting the reconstruction algorithm, prerequisite sections are listed
first --- Section \ref{sect:operators} and \ref{sect:lossfunc}. The main
algorithmic routine is then thoroughly discussed in Section \ref{sect:algorithm}.

\subsection{Beam manipulation operators}{\label{sect:operators}}
The algorithm slowly ``steers'' a randomly generated Gaussian beam toward its
``true'' position and shape. This transformation is accomplished through several
operators, which modify the position $(x,y)$ of each particle in the beam. The
transformations are made in small steps ( $\delta_{x,y}$) for better accuracy
and stability. The list of the operators are as follows:
\begin{itemize}
	\item \textit{Translation}: This operator is required for finding the correct beam position.
	\begin{eqnarray}
		\label{eq:translation_}
		(x,y)\longrightarrow (x \pm \delta_x , \quad y),\\
		\label{eq:translation}
		(x,y) \longrightarrow (x, \quad y \pm \delta_y).
	\end{eqnarray}

	\item \textit{Scaling}: The beam size and shape change by stretching the beam with respect to its center.
	\begin{eqnarray}
		(x,y)\longrightarrow (x \cdot (1\pm \delta_x) , \quad y), \\
		(x,y) \longrightarrow (x, \quad y \cdot (1\pm \delta_y)).
		\end{eqnarray}

	\item \textit{Shearing}: This operator introduces a coupling between the $x$ and $y$ distributions, hence a rotation.
	\begin{eqnarray}
		(x,y)\longrightarrow (x \cdot (1\pm y\delta_x) , \quad y), \\
		(x,y) \longrightarrow (x,\quad  y \cdot (1\pm x\delta_y)).
	\end{eqnarray}
	
	\item \textit{Bending}: This operator helps achieving a variety of distortions.
	\begin{eqnarray}
		(x,y)\longrightarrow (x \cdot (1\pm \textnormal{abs}(y)\cdot \delta_x) , \quad y), \\
		(x,y) \longrightarrow (x, \quad y \cdot (1\pm \textnormal{abs}(x)\cdot \delta_y)).
		\label{eq:bending}
	\end{eqnarray}

\end{itemize}

As an example, by means of Equations \ref{eq:translation_} and
\ref{eq:translation}, a test beam \textbf{X$^\textnormal{try}$} moves by
$(\pm\delta_x, \pm \delta_y)$, chasing the minimum value of the loss function,
and settles where it minimizes. The method is heavily inspired by the gradient
descent \cite{ref:gr_desc} and simulated annealing \cite{ref:sim_anneal}
algorithms. All the operations are applied in parallel, and the best operation
is selected according to the resulting loss function. The resulting state of the
beam profile most likely sits at a local minimum. As mentioned above, we are
using the genetic algorithm to introduce a randomness and an escape mechanism
from the local minima. 
	
It is clear that this is not the ultimate list of operators for beam
transformations. For instance, an ``S-shaped'' beam cannot be produced by them
from an initially Gaussian beam. This limitation does not appear in our study
because both the test beam profiles (\textbf{X$^\textnormal{true}$}) and the
reconstructed profiles (\textbf{X$^\textnormal{try}$}) are generated by
modifying a Gaussian beam by the same operators. However, this does not
invalidate the conclusions of this study as long as one can define operators
relevant to a particular shape. 

Finally, as mentioned above, the ambiguity from $\sigma_{x}^2 - \sigma_{y}^2$
also limits the performance of the algorithm. Therefore, we introduced beam size
knowledge at the loss function to favor solutions that are compatible with a
presumed beam size. 

\subsection{Loss function}{\label{sect:lossfunc}}
The search algorithm accepts or rejects a transformation according to its loss
function. Choice of a loss function is somewhat an arbitrary procedure, and can
affect the performance significantly. It is composed of two parts in this
routine.

The first part is the mean square error (MSE) of the signal that is induced at the electrodes:
\begin{equation}
	\textnormal{MSE}(\textbf{Y}^\textnormal{try}, \textbf{Y}^\textnormal{true})= \frac{1}{N_e}(\textbf{Y}^\textnormal{try}- \textbf{Y}^\textnormal{true})(\textbf{Y}^\textnormal{try}- \textbf{Y}^\textnormal{true})^\textbf{T},
	\label{eq:mse}
\end{equation}
where $N_e$ is the number of electrodes, \textbf{Y}$^\textnormal{true}$ and
\textbf{Y$^\textnormal{try}$} represent the true signal and the signal obtained
after transformation, respectively, and \textbf{T} is the transpose of the
matrix. 

The degeneracy problem requires a knowledge of the beam dimensions, $\langle x^2
\rangle$ and $\langle y^2 \rangle$. The second part of the loss function,
so-called exponential penalty (EP), is given as
\begin{equation}
\begin{split}
	\Delta_x &\equiv \left(\sqrt{\langle x^2 \rangle ^\textnormal{try}} - \sqrt{\langle x^2 \rangle}\right)^2, \\
	\Delta_y &\equiv \left(\sqrt{\langle y^2 \rangle ^\textnormal{try}} - \sqrt{\langle y^2 \rangle}\right)^2, \\
	\textnormal{EP}(\textbf{Y}^\textnormal{try}, \textbf{Y}^\textnormal{true})&=\exp(w\Delta_x) + \exp (w\Delta _y),
\end{split}
\end{equation}
where $w$ is practically the relative weight of the EP term. Then, the combined
loss function becomes
\begin{equation}
L(\textbf{Y}^\textnormal{try}, \textbf{Y}^\textnormal{true})=\textnormal{MSE}(\textbf{Y}^\textnormal{try}, \textbf{Y}^\textnormal{true}) + \textnormal{EP}(\textbf{Y}^\textnormal{try}, \textbf{Y}^\textnormal{true}),
\label{eq:loss}
\end{equation}
The EP favors the solutions that have a size similar to the given beam size. If
the loss function is purely EP, then the algorithm converges to a Gaussian beam
with the correct beam size, but loses the profile and position information. In
contrast, if it is purely MSE, then the algorithm likely converges to one of the
degenerate solutions with an arbitrary size. According to our tests, $w \approx
10^4$ yields acceptable results for a wide range of beam profiles. However, with
a prior knowledge of the approximate beam profile or the beam size, it can be
readjusted to boost the performance. It can also be calibrated by independent
measurements of other BPMs as well. It is worth emphasizing again that the
choice of the operators, $\pm \delta_{x,y}$, and $w$ is rather arbitrary and
should vary at different applications.

Another approach is to generate the reproduction candidates solely via the MSE
and eventually filter them by the EP according to the beam size. However, this
approach suffers from unnecessary steering of the beam profile toward unlikely
$\langle x^2 \rangle$ and $\langle y^2 \rangle$ values during the minimization
process. We found it computationally beneficial to embed the EP into the loss
function.

\subsection{Algorithm}
\begin{figure*}
\centering
\includegraphics[width=0.8\linewidth]{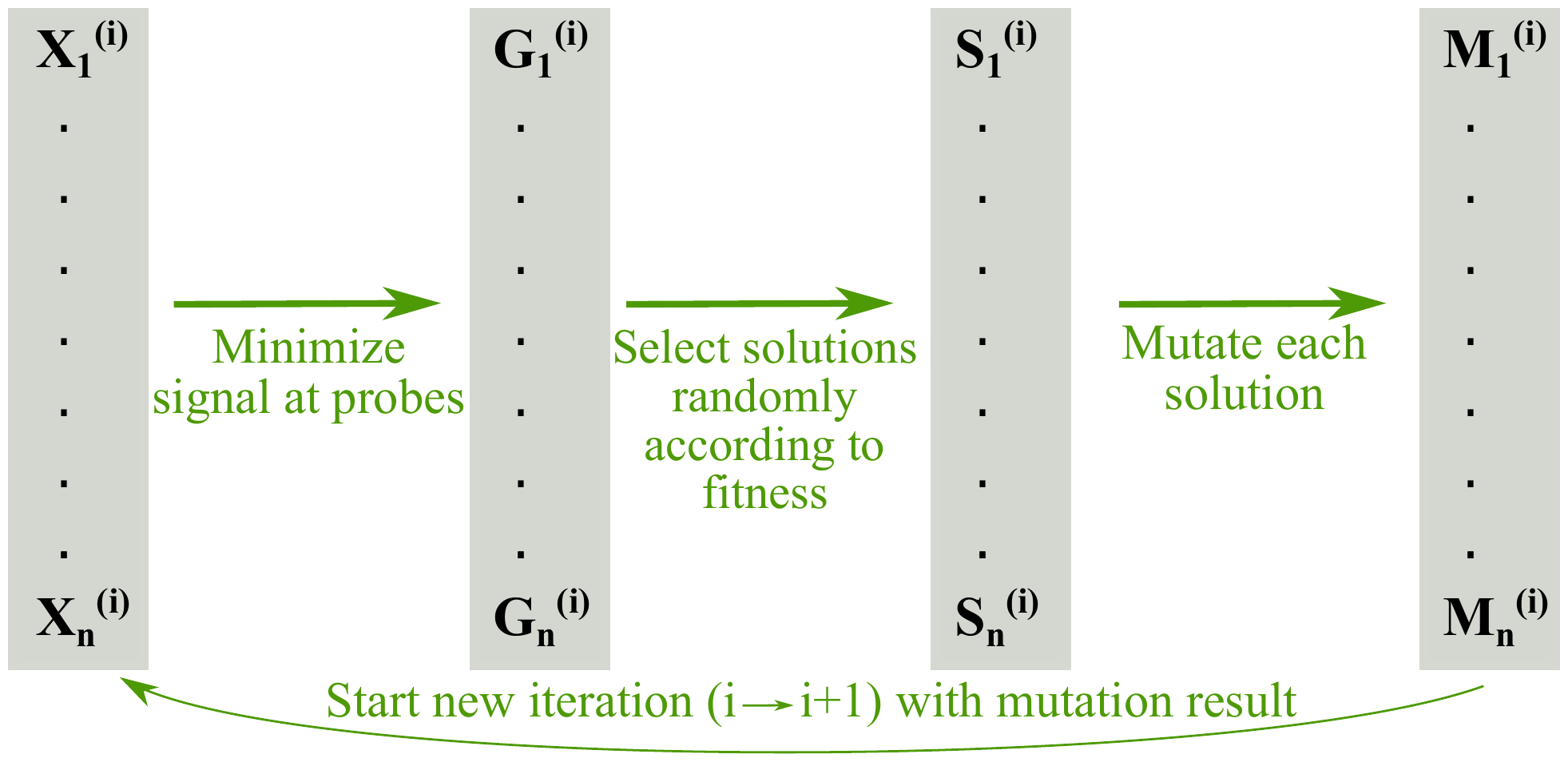}
\caption{The reconstruction routine is composed of several layers of operations. Firstly, $n$ random beams are initialized and then transformed by a gradient descent-like method to reach minima, either local or global. Then, some solutions are selected according to their loss functions, and mutated for the next iteration.}
\label{fig:bpm_algorithm}
\end{figure*}

{\label{sect:algorithm}}
The reconstruction routine is depicted in Figure \ref{fig:bpm_algorithm}.
Initially, a random beam (\textbf{X$^\textnormal{true}$}) with an arbitrary
distortion is created and the corresponding \textbf{Y$^{\textnormal{true}}$} is
calculated for reference. In a real BPM measurement, only
\textbf{Y$^{\textnormal{true}}$} will be known. Then, a minimization algorithm
starts with $n=50$ candidate Gaussian beams of random shapes and locations,
represented with \textbf{X$_{1\rightarrow n}$}. Each beam induces signals on the
probes, shown with the vector \textbf{Y$^{\textnormal{try}}$}. The beams are
transformed in parallel by several operators one-by-one to achieve signals as
close as possible to the given \textbf{Y$^{\textnormal{true}}$}. The above
procedures do not guarantee the correct beam profile because it may be stuck at
a local minimum. Then, according to their loss function values, $n$ solutions
are selected for mutation. In contrast to the cross breeding of the genetic
algorithm \cite{ref:gen_alg}, the next generation beams are selected randomly,
with probabilities inversely proportional to the loss function. This is also
known as ``roulette wheel selection'' or ``fitness proportionate selection''.
Note that in this selection, some solutions $G_i$ may be selected multiple times
and some others may not be selected at all. Then, the selected candidates are
mutated by randomly applying one of the operators of Section
\ref{sect:operators}, only with a higher amplitude. The last step ensures that
the candidates are steered away from the local minima. This completes the
iteration and the $n$ newly generated beam profiles are returned as starting
candidates for the next iteration. Algorithm \ref{alg:algorithm} summarizes the
process in pseudo code.

%\clearpage
%\begin{strip}
\removelatexerror
\begin{algorithm*}[H]
	% \SetAlgoLined
	\DontPrintSemicolon
	\KwIn{\textbf{Y$^{\textnormal{true}}$}}
	\KwOut{\textbf{X$^{\textnormal{try}}$}}
	{Initialize $n$ parallel instances of randomly generated beams \textbf{X$_{1 \rightarrow n}$}}\;
	\For{iterations}
	{
		\ForEach{\textbf{X$_{1 \rightarrow n}$}}{
			\tcc{Search for local minimum}
			\Repeat{$L(\mathbf{Y}_{1 \rightarrow n}^\textnormal{try}, \mathbf{Y^{\textnormal{true}}}) < L(\mathbf{Y}_{1 \rightarrow n}, \mathbf{Y^{\textnormal{true}}})$ is found} {
				$\mathbf{Y}_{1 \rightarrow n} = J_{\textnormal{image}}(\mathbf{X}_{1 \rightarrow n})$ from Equation \ref{eq:current}\;
				$\mathbf{X}_{1 \rightarrow n}^\textnormal{try}$ = [$f(\mathbf{X}_{1 \rightarrow n})$ for $f$ in operators specified in Section \ref{sect:operators}] \tcc {Transform the beams}
				$\mathbf{Y}_{1 \rightarrow n}^\textnormal{try} = J_{\textnormal{image}}(\mathbf{X}_{1 \rightarrow n}^\textnormal{try})$ \tcc {Electrode signals after transformation}
			} % end of repeat
			 \tcc{$L$ is the loss function, defined in Section \ref{sect:lossfunc}}
			$\mathbf{G}_{1 \rightarrow n} =  \mathbf{X}_{1 \rightarrow n}^\textnormal{try}$ \;
		}
		\tcc{At this point, every $\mathbf{X}_{1 \rightarrow n}$ is transformed to a $\mathbf{G}_{1 \rightarrow n}$ that represents a minimum}
		Select $n$ solutions \textbf{S$_{1 \rightarrow n}$} from \textbf{G$_{1 \rightarrow n}$} \;
		Mutate $\mathbf{S}_{1 \rightarrow n}$ for the next iteration:  $\mathbf{S}_{1 \rightarrow n}  \longrightarrow \mathbf{X}_{1 \rightarrow n}$\;
	}
	\caption{Initially, a test beam \textbf{X$^{\textnormal{true}}$} is created and the corresponding signal \textbf{Y$^{\textnormal{true}}$} on the electrodes is calculated. The algorithm aims to reconstruct \textbf{X$^{\textnormal{true}}$} from a set of randomly generated beams, by transforming them in parallel, which eventually induce \textbf{Y$^{\textnormal{true}}$}. It makes use of a genetic algorithm with a gradient descent-like minimization. A schematic view of the algorithm is given in Figure \ref{fig:bpm_algorithm}.}\label{alg:algorithm}
\end{algorithm*}
%\end{strip}

We attempted mapping $\textbf{X}$ onto $\textbf{Y}$ by means of data-driven deep
neural networks as well (More about this in \ref{sec:neuralnetwork}).
Such a method, however, biases the predictions toward previously seen data and
is unable to predict unencountered beam profile shapes.  Eventually, we found
the performance of the presented method to be superior not only because it
requires no precursory training, but also it produces more robust results, free
of possible neural network artifacts.

\section{Reconstruction tests}{\label{sect:results}}
For every test, we determined a ``true'' beam of random shape and position,
which had a ``true'' signal on the electrodes. Then, we followed the algorithm
that was explained in Section \ref{sect:method} for reconstruction. At the end,
comparison between the true and reconstructed beams
($\textbf{X}^\textnormal{true}$ and \textbf{X}$^\textnormal{rec}$, respectively)
is made by using Structural Similarity Index Measure (SSIM), which is defined as
\begin{equation}
	  \textnormal{SSIM}(\textbf{X}^\textnormal{true},\textbf{X}^\textnormal{rec}) = \frac{(2\mu_\textnormal{t}\mu_\textnormal{r} + C_1)(2\sigma_\textnormal{tr} + C_2)}{(\mu_\textnormal{t}^2 + \mu_\textnormal{r}^2 + C_1)(\sigma_\textnormal{t}^2 + \sigma_\textnormal{r}^2 + C_2)}.
	  \label{eq:ssim}
\end{equation}
Here, $\mu$ and $\sigma$ are the mean and the standard deviation of the pixel
brightness, subscripts `t' and `r' represent ``true'' and ``reconstructed''
respectively. $\sigma_\textnormal{tr}$ refers to covariance of `t' and `r'.
$C_1$ and $C_2$ are normalization constants. The SSIM score ranges from -1.0 to
1.0, with the latter corresponding to an identical copy. 

Rather than calculating Equation \ref{eq:ssim} once for the whole images, they
are split into sub-regions and the calculations are made at that scale. Then,
the SSIM scores are averaged over the individual sub-region pairs, weighted with
a circular Gaussian filter of $1.5\sigma$. The SSIM is preferable for a 2D image
comparison since it mimics human preception. More detailed discussion on SSIM,
selection of the coefficients and the sub-regions, and a comparison with MSE of
Equation \ref{eq:mse} can be found at References \cite{ref:ssim1} and
\cite{ref:ssim2}.

\begin{figure*}
	\includegraphics[width=\linewidth]{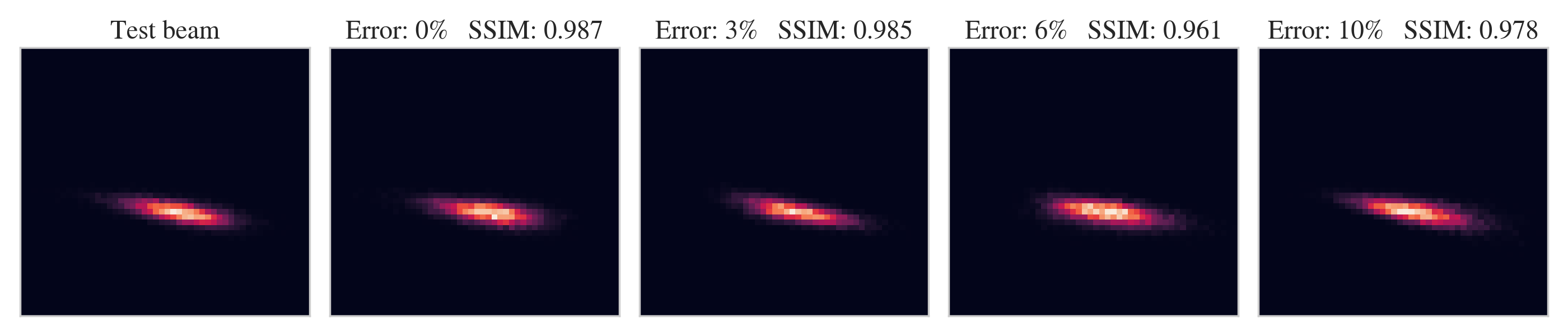}
	\includegraphics[width=\linewidth]{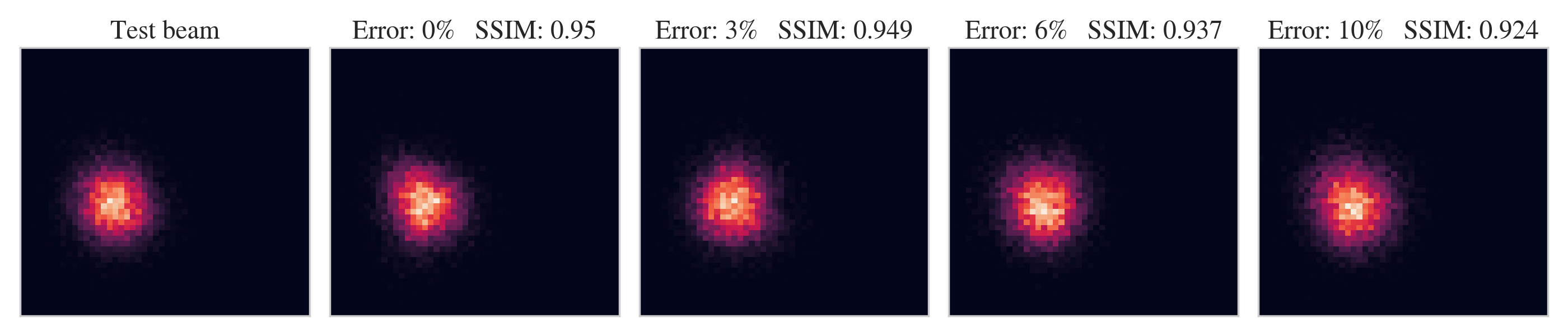}
	\includegraphics[width=\linewidth]{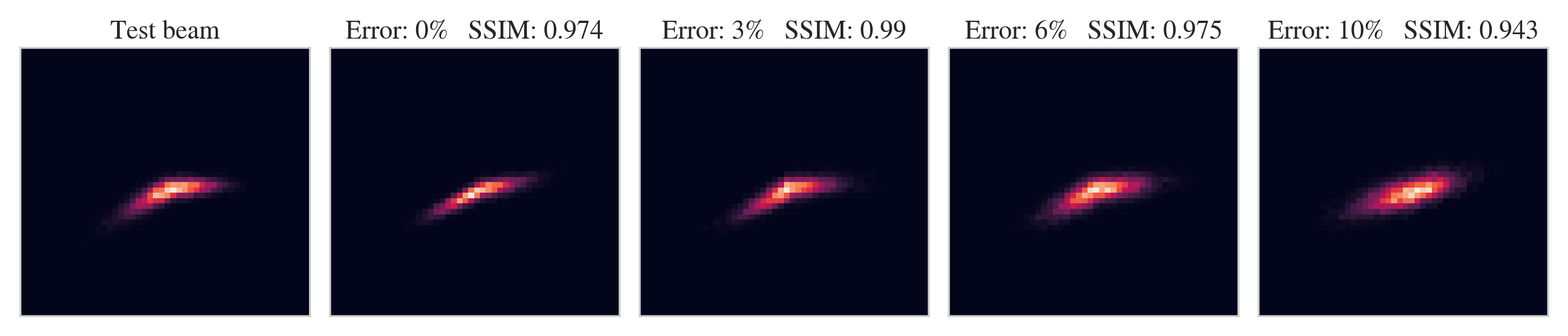}
	\includegraphics[width=\linewidth]{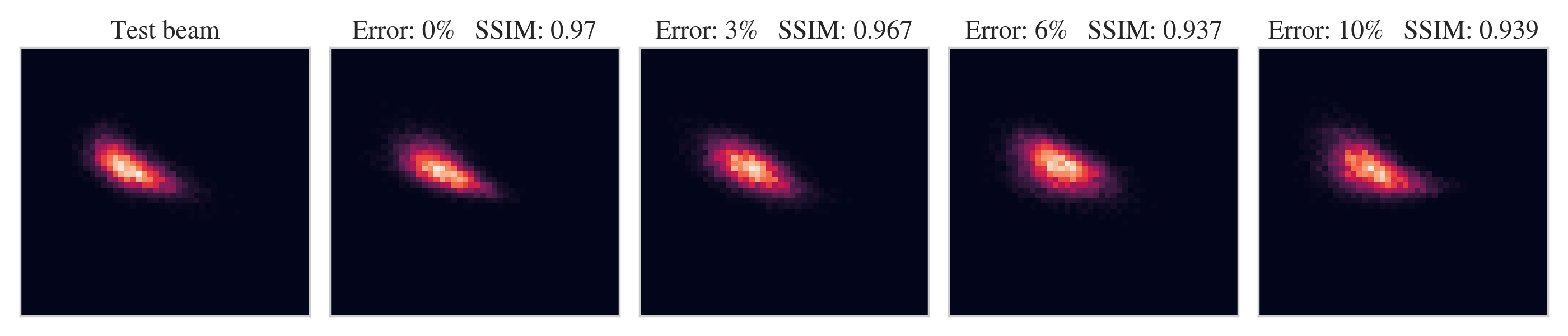}
	\includegraphics[width=\linewidth]{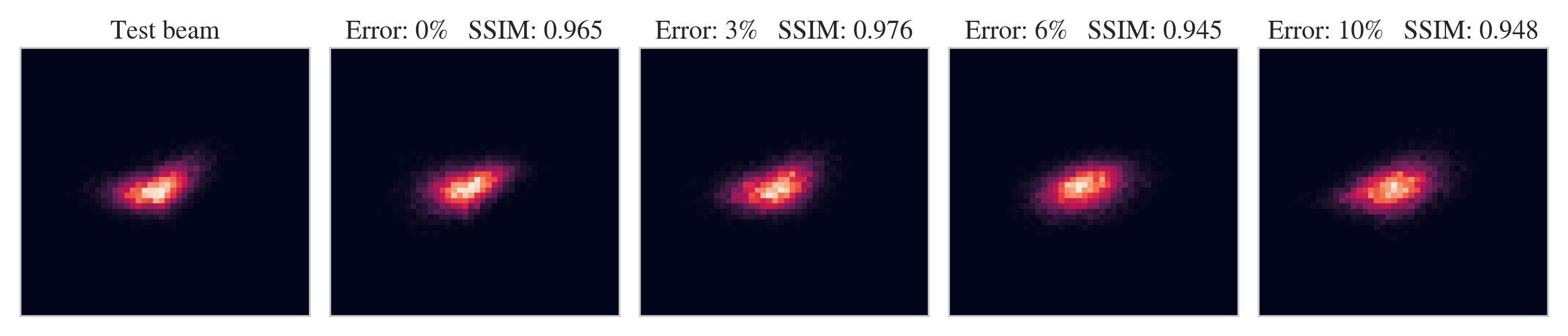}
	\caption{Test beam profiles and the reconstructed beam profiles for different cases. Location of the probes and the scale of the images are identical with Figure \ref{fig:transform}. Each row corresponds to a separate test beam, given in the first column. The columns on the right correspond to beams with size errors ranging between 0\% and 10\%. SSIM of each solution, compared to the test beam is shown above the images as reference.  Figure \ref{fig:ssim_vs_noise} shows the overall SSIM results for a larger ensemble.}
	\label{fig:test_results}
\end{figure*}

As mentioned in Section \ref{sect:lossfunc}, the beam size knowledge plays an
important role in the reconstruction routine. For every test, we have also
applied the reconstruction algorithm with incorrect beam size values to see how
the performance degrades. Figure \ref{fig:test_results} shows a variety of test
beams (the leftmost column) and the corresponding reconstructions with
increasing errors in the beam size knowledge at every column. The beam size
errors correspond to $\sigma_x$ and $\sigma_y$ at the same time, because the
reconstruction error becomes less if one of them has a smaller error.

\begin{figure}
\centering
\includegraphics[width=0.5\linewidth]{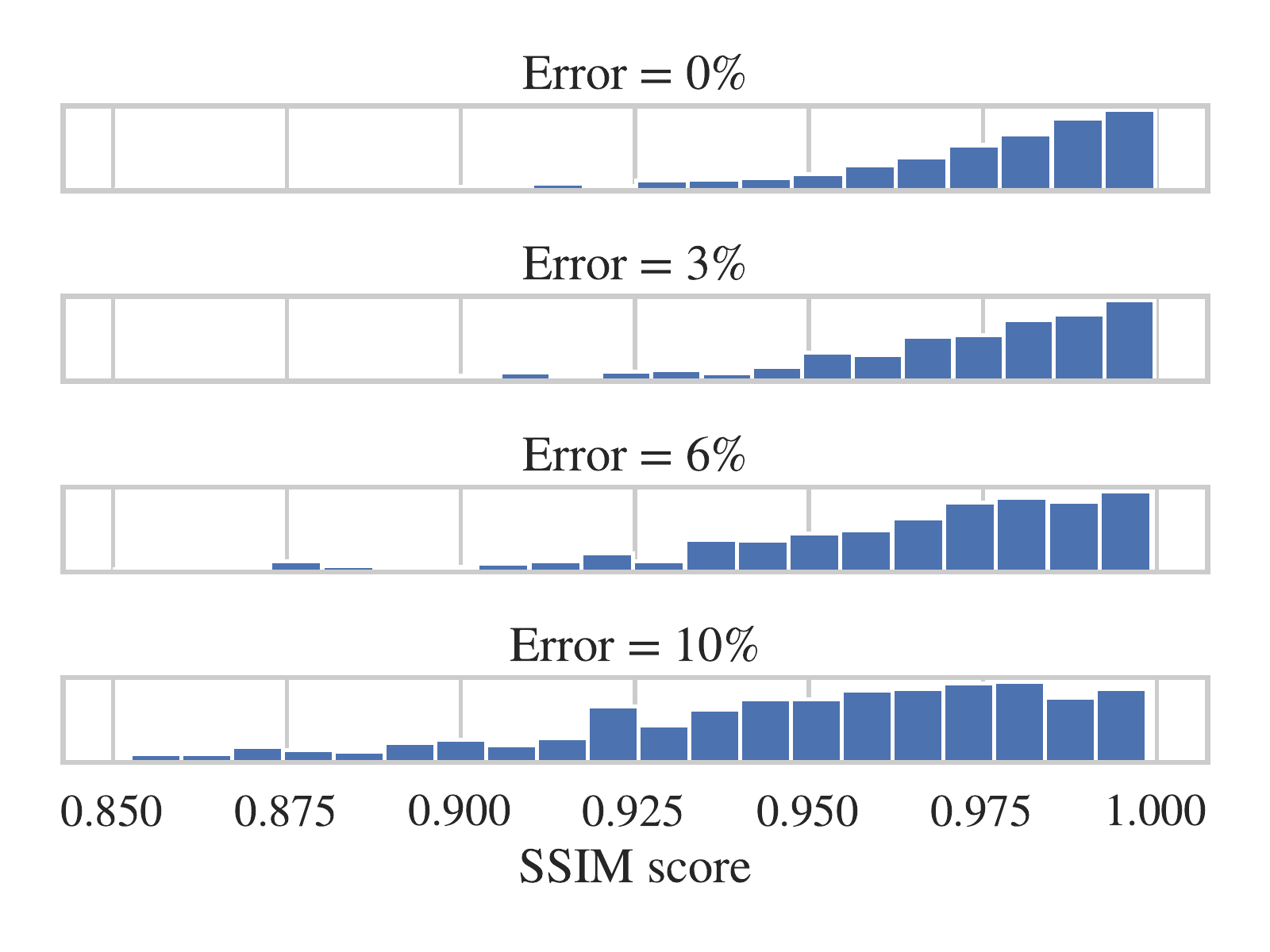}
\caption{Histogram of SSIM vs. beam size error, which is obtained from 500 test beams. The method performs effectively when the beam size is known well. The tail is likely to originate from the extremely distorted beams, which is always harder to reconstruct. The method yields acceptable results up to 10\% error, while it performs quite better at lower errors.} 
\label{fig:ssim_vs_noise}
\end{figure}

Elongated beams have a clear quadrupole moment, which is relatively easy to pick
up by the electrodes. Therefore, they can be perfectly reconstructed if the beam
size is known well (Row 1). With an increase in the beam size error, the beam
becomes thicker while $\sigma_x^2 - \sigma_y^2$ remains the same. This effect
becomes more visible with beam size errors larger than 20\%. The signal at the
electrodes do not vary much for round beams (Row 2). Therefore, the algorithm
works very well if the beam size is known well. Otherwise, the beam shape is
protected except for very small beams, while its size changes in proportion to
the beam size error. This effect is also barely visible for 10\% level beam size
error. While distorted beams (as shown in Rows 3, 4, and 5) can also be
reconstructed quite precisely, there are occasional cases that banana shapes are
interpreted as triangular, and vice versa. This originates from the similarity
of the probe signals for each case. It is worth noting that <10\% beam size
error, where the BPM performs quite well for all studied profiles, is reasonable
for the proton EDM experiment. More details on this are given in the
\ref{sec:beamsizeappendix}.  Figure \ref{fig:ssim_vs_noise} shows the SSIM
comparison between the test beam and the reconstructed beams for an ensemble of
500 particles with random profiles. Comparison with Figure
\ref{fig:test_results} shows that a SSIM score around 0.95 and beyond usually
indicates a perfect match, which is likely to be achieved with good knowledge of
the beam size. For the 10\% beam size error, majority of the reconstructed
profiles have SSIM scores better than 0.90. The results get significantly better
at 6\% beam size error and below. 

\begin{figure}
\centering
\includegraphics[width=0.5\linewidth]{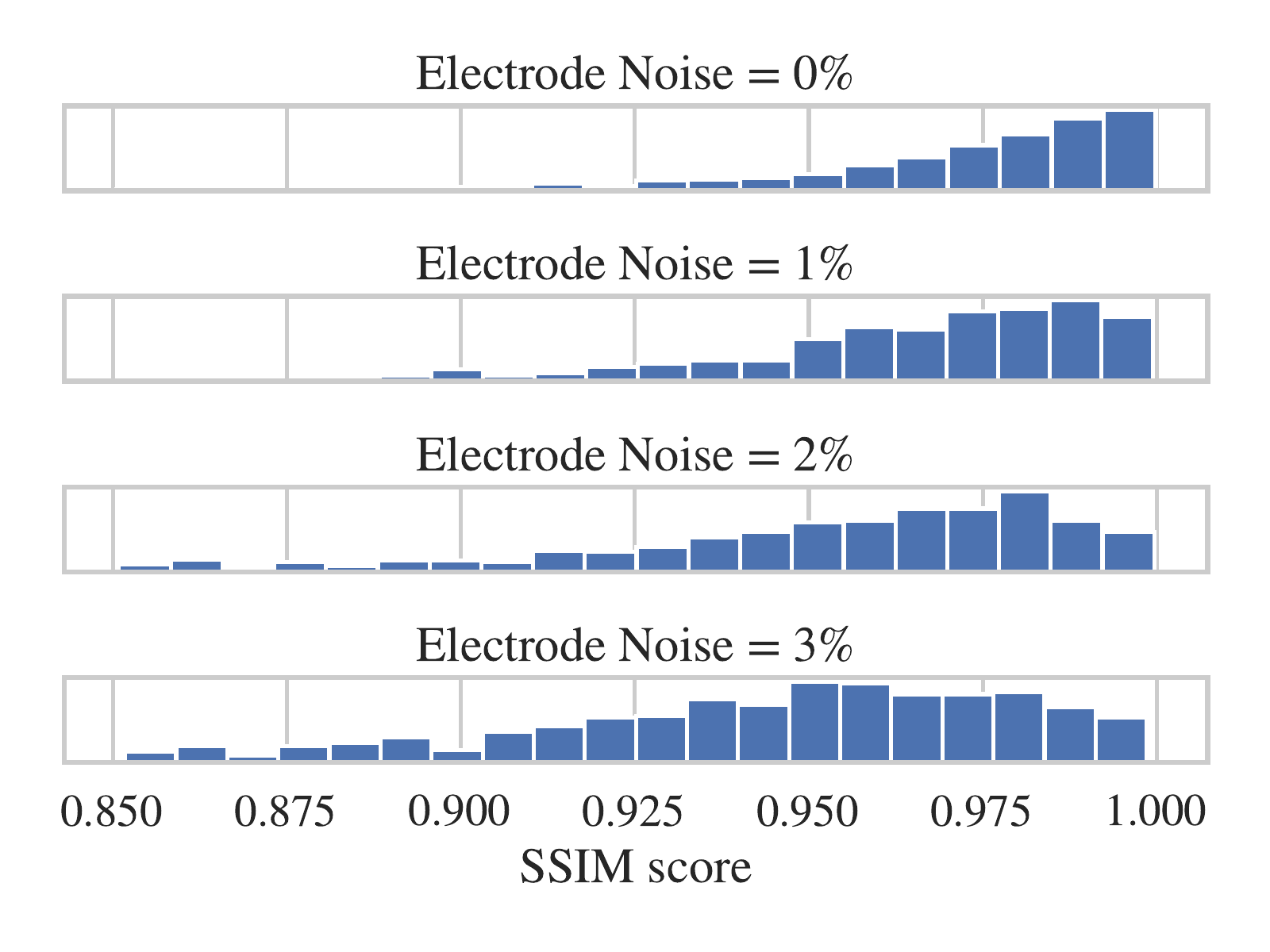}
\caption{Histogram of SSIM vs. random rms $\sigma$ probe noise, which is
  obtained from 500 test beams. Random rms $\sigma=$0--3\% noise is applied on each of the 8
  probes of the profile monitor. The quality of the results (SSIM score) does
  not seem to degrade significantly.}
\label{fig:probe_noise}
\end{figure}

Finally, as confirmed by additional
simulations, the SSIM score is not seriously affected by up to $\pm 3\%$ random
probe noise --- Figure \ref{fig:probe_noise}. Such random probe noise is added
to imitate real life total noise levels. It is equally important how such noise
translates to voltage levels in a realistic physical design. Assuming that the thermal 
noise contribution would be around,
\[v_{n} = \sqrt{4 k_B T R \Delta f} =  \sqrt{4 \cdot 1.38 \cdot 10^{-23}~\mathrm{J}/\mathrm{K} \cdot 300~\mathrm{K} \cdot 50 \Omega \cdot 100~\mathrm{MHz}}=9~\mu \textrm{V}.\]
$50~\Omega$ corresponds to pickup grounding impedance and $100~\mathrm{MHz}$ corresponds to expected (conservative) bandwidth. 
The voltage signal ($I_{\textnormal{beam}} \cdot Z_{\mathrm{transfer}}$) including the aforementioned acceptable 3\% noise would be,
\[ 1.7~\mathrm{mA} \cdot Z_{\mathrm{transfer}} \cdot 0.03 = 51~\mu \textrm{V},\]
where the transfer impedance was assumed to be $Z_{\mathrm{transfer}}=1\Omega$ as a reference. In such a configuration, the expected power signal-to-noise ratio is 31. As the bunches 1-to-80 are supposed to be similar at each turn, averaging over those bunches can be considered for further improvement.

\section{Summary and conclusion}

We presented a novel non-destructive beam profile visualization technique that
requires a button beam position monitoring hardware and machine learning
enhanced data processing. Genetic algorithm starts off  with an array of  first
generation random beam profiles which are then transformed by successive
application of transformations (given in Section \ref{sect:operators}) toward
the minimal loss function (Equation \ref{eq:loss}). Once the entire generation
of beam profiles reach their respective local minima, several beam profiles are
selected randomly with probabilities inversely proportional to their loss
functions, reproduced, and then mutated toward use in the next generation. This
process repeats for a predetermined number of generations. The major limitation
is the degeneracy due to $\sigma_x^2-\sigma_y^2$. In principle, it can be
eliminated with a perfect knowledge of the beam size. In the proton EDM
experiment, the error in beam size estimation is expected to be known within a
few percent because of continuous extraction. Depending on the application,
additional beam transformation operators that are more compatible with the
presumed beam profiles can be implemented. A simplified code and an example are
given at Refs \cite{ref:code} and \cite{ref:example}, respectively.

\section {Acknowledgments}
This work was supported by IBS-R017-D1 of the Republic of Korea. We would like
to thank Yannis K. Semertzidis for helpful discussions. We would also like to
thank Changkyu Sung for pointing the degeneracy problem due to the quadrupole
moment, and Sanzhar Bakhtiyarov and Adil Karjauv for helpful discussions on
machine learning algorithms.

\appendix
\section {Beam size and effect of the $\beta$ and $\epsilon$ drifts}\label{sec:beamsizeappendix}

\begin{figure}
\centering
\includegraphics[width=0.7\linewidth]{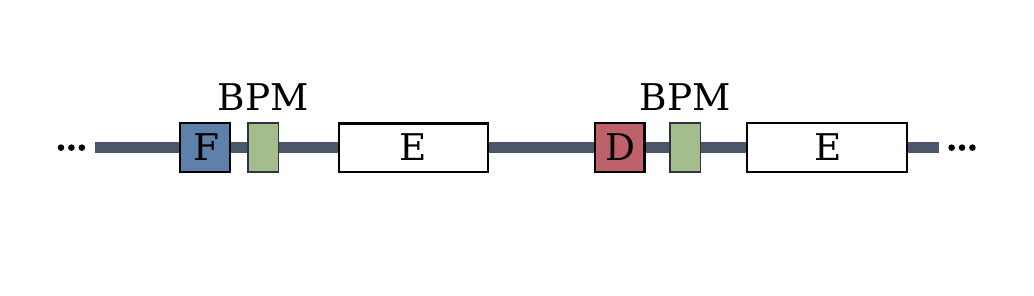}
\caption{A FODO cell with electric deflectors (E), focusing (F) and defocusing (D) quadrupoles, BPM's, and straight sections (gray lines). Sextupoles overlap with quadrupoles.}
\label{fig:fodo}
\end{figure}

\begin{figure}
\centering
\includegraphics[width=0.5\linewidth]{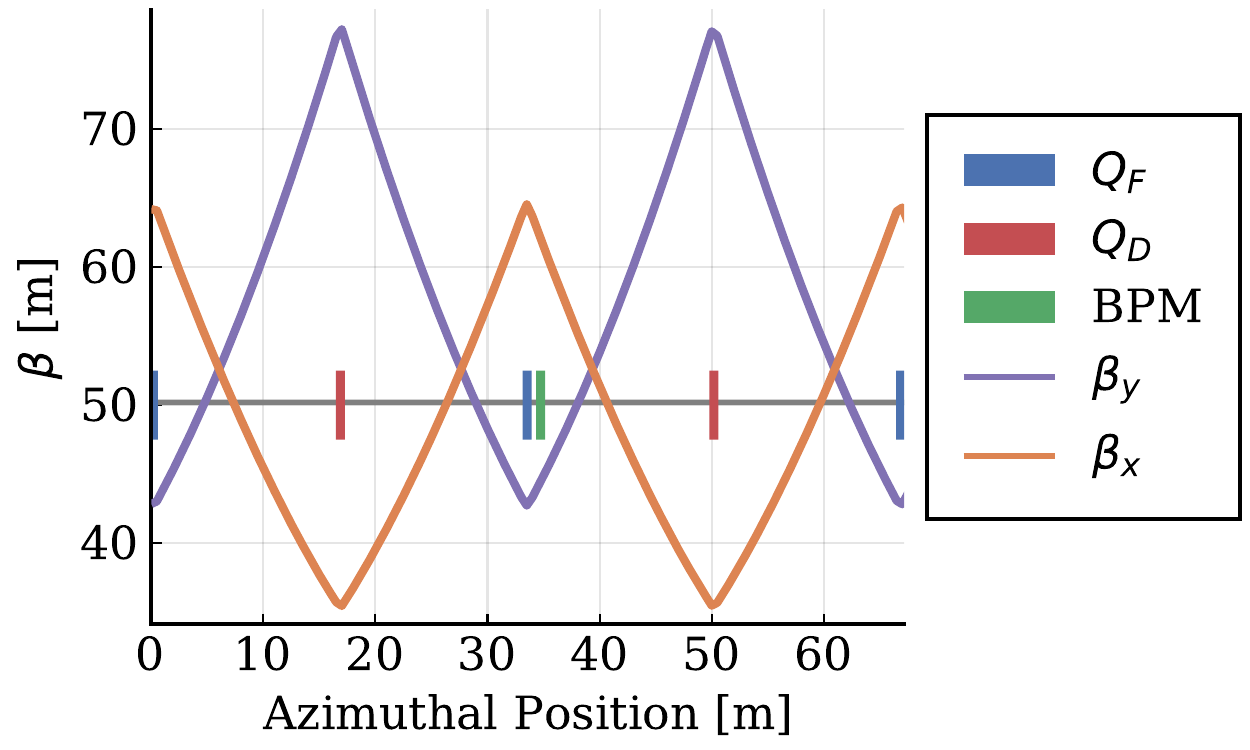}
\caption{Beta functions in two cells. The focusing and defocusing quadrupoles are shown with blue and red vertical lines. The BPM (shown in green) is located 1 m away from the quadrupole.}
\label{fig:beta}
\end{figure}

The pEDM storage ring is composed of 24 cells, each having electric deflectors,
drifts, sextupoles, and focusing and defocusing magnetic quadrupoles as depicted
in Figure \ref{fig:fodo}. The beta functions are symmetric at every cell as
shown in Figure \ref{fig:beta}, which covers two cells with the quadrupoles and
the BPM indicated. The emittances of the beam are $\epsilon_x=0.214$ mm-mrad and
$\epsilon_y=0.25$ mm-mrad. The beam will be continuously extracted during
storage at the polarimeter target. Therefore, its size will be fixed at that location. Upon beta function measurements, it can also be estimated at other locations.

Some storages in the experiment will periodically be dedicated to beam and field
diagnostics, and field corrections, instead of EDM measurements. Beta functions
can as well be determined at those diagnostic storages. According to our
studies, the emittance growth time is roughly 40 minutes in all directions.
Combined with continuous extraction, this guarantees a fixed beam size at the
polarimeter location during storage ($\approx 20$ minutes). However, the
stability of the quadrupoles in long term may be a concern as it has an effect
on the beam size. This is investigated in two ways by i) changing the focusing
strength of the neighbor quadrupole, and ii) randomly changing the focusing
strength of all the quadrupoles. Note that these scenarios refer to quadrupole
strength variations at the EDM measurement storages. 

%As explained above, our reconstruction method relies on finding a beam distribution that generates a signal as close as possible to the electrode measurements. Like all minimization algorithms, it can fail because of local minima. Moreover, for a Gaussian beam, there may be more than one global minima, determined by $\sigma^2 \equiv \sigma_x^2-\sigma_y^2$ and  $\bar{x}^2-\bar{y}^2$ values. Restricting the search by an already known beam size (represented by $\sigma_x$ and $\sigma_y$ of presumed Gaussian distributions on each axis) improves the solution significantly.

The RMS beam size is roughly determined by the emittance $\epsilon$ and the beta
function $\beta$ as
\begin{equation*}
\begin{split}
\sigma_x^2\propto \langle x^2 \rangle & = \epsilon_{x} \beta_x, \\
\sigma_y^2\propto \langle y^2 \rangle & = \epsilon_{y} \beta_y,
\end{split}
\end{equation*}
where $x$ and $y$ are the horizontal and vertical positions with respect to the
beam center, respectively. The brackets represent average over particles in the
beam, and the subscripts represent the horizontal or vertical directions.  The
ratio of the horizontal beam size to the vertical is
\begin{equation*}
\frac{\sigma_x^2}{\sigma_y^2} = R_\epsilon R_\beta,
\end{equation*}
where $R_\epsilon \equiv\epsilon_x/\epsilon_y$ and $R_\beta \equiv \beta_x/\beta_y$. Then, one gets
\begin{eqnarray*}
	\sigma^2 =&\left(   R_\epsilon R_\beta-1 \right) \sigma^2_y, \\
	\sigma^2_y=&\frac{\sigma^2}{R_\epsilon R_\beta -1}, \\
	\sigma^2_x =&\sigma^2 \left( 1+ \frac{1}{R_\epsilon R_\beta -1} \right).
\end{eqnarray*}
$\sigma^2$ is obtained by the BPM, and $R_\epsilon$ and $R_\beta$ are given
either by analytical estimations or independent measurements. The error in
$\sigma_x^2$ and $\sigma_y^2$ is
\begin{equation}
\delta \sigma^2_{x,y}=-\frac{R_\epsilon \delta R_\beta+R_\beta \delta R_\epsilon}{(R_\epsilon R_\beta-1)^2}\sigma^2, 
\label{eq:error}
\end{equation}
where $\delta R_\epsilon$ and $\delta R_\beta$ are the errors of $R_\epsilon$
and $R_\beta$ estimations, respectively. $\delta R_\beta$ is defined as
\begin{equation}
\delta R_\beta = \frac{\delta \beta_x}{\beta_y}-\frac{\beta_x\delta \beta_y}{\beta_y^2},
\label{eq:err_on_R}
\end{equation}
where $\delta \beta_{x,y}$ are the errors on the beta functions. Setting $\delta
\beta_{x}=b_{x}\beta_{x}$ and $\delta \beta_{y}=b_{y}\beta_{y}$, Equation
\ref{eq:err_on_R} simplifies to $\delta R_\beta=(b_x-b_y)R_\beta$.  Similarly,
by using  $\delta \epsilon_{x}=a_{x}\epsilon_{x}$ and $\delta
\epsilon_{y}=a_{y}\epsilon_{y}$, one gets $\delta R_\epsilon =
(a_x-a_y)R_\epsilon$. Inserting $\delta R_\beta$ and $\delta R_\epsilon$ into
Equation \ref{eq:error}, one obtains 
\begin{equation}
\delta \sigma^2_{x,y}= - \frac{R_\epsilon R_\beta (a_x-a_y+b_x-b_y)}{(R_\epsilon R_\beta-1)^2}\sigma^2.
\label{eq:error_final}
\end{equation}

\begin{figure}
	\centering
	\includegraphics[width=0.6\linewidth]{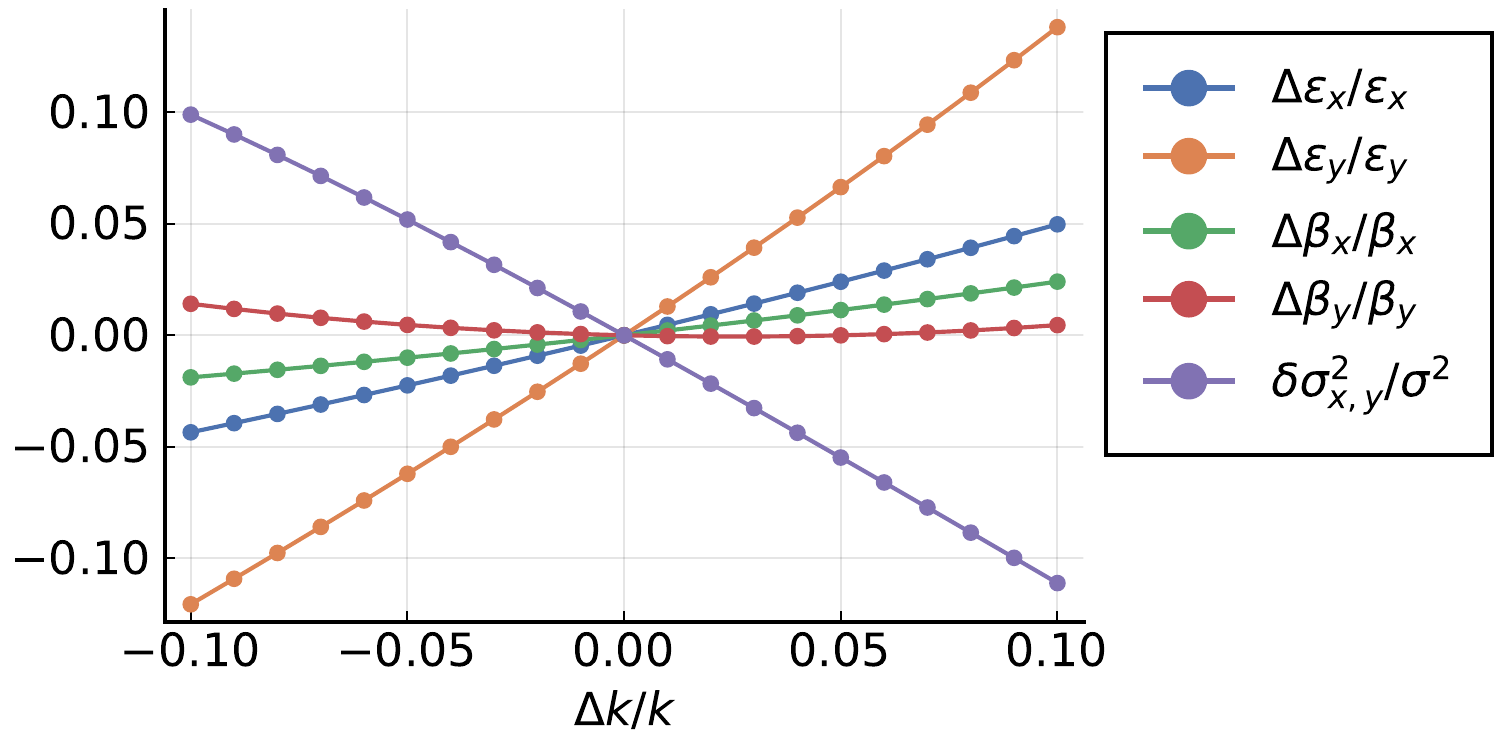}
	\caption{The change in the beta functions, emittance, and the beam size as the quadrupole strength next to the BPM is varied.}
	\label{fig:next_quad_offset}
\end{figure}

For the lattice proposed at Ref \cite{ref:symm_hyb_ring}, $R_\epsilon\approx
0.9$ and $R_\beta$ is 1.5 or 0.47, depending on the BPM location. As a result,
one half cell is roughly an order of magnitude more sensitive than the other to
the emittance and beta function drifts. Figure \ref{fig:next_quad_offset} shows
the variation of the emittances, beta functions, and the beam size (from
Equation \ref{eq:error_final}) for the ``worse'' half cell as a function of
variations of the neighboring quadrupole strength ($\Delta k/k$). The data is
obtained by one-particle beam dynamics simulations inside the storage ring of
Ref \cite{ref:symm_hyb_ring}. Beam stability and resonance issues are not
considered in this treatment. The change in the beam size is almost linearly
proportional to $\Delta k/k$, and it is at the same order of the changes in the
beta functions and the emittances. 10\% change in $\Delta k/k$, resulting in
10\% change in the beam size, is quite tolerable for the BPM as shown in Figures
\ref{fig:test_results} and \ref{fig:ssim_vs_noise}.  

\begin{figure}
	\centering
	\includegraphics[width=0.5\linewidth]{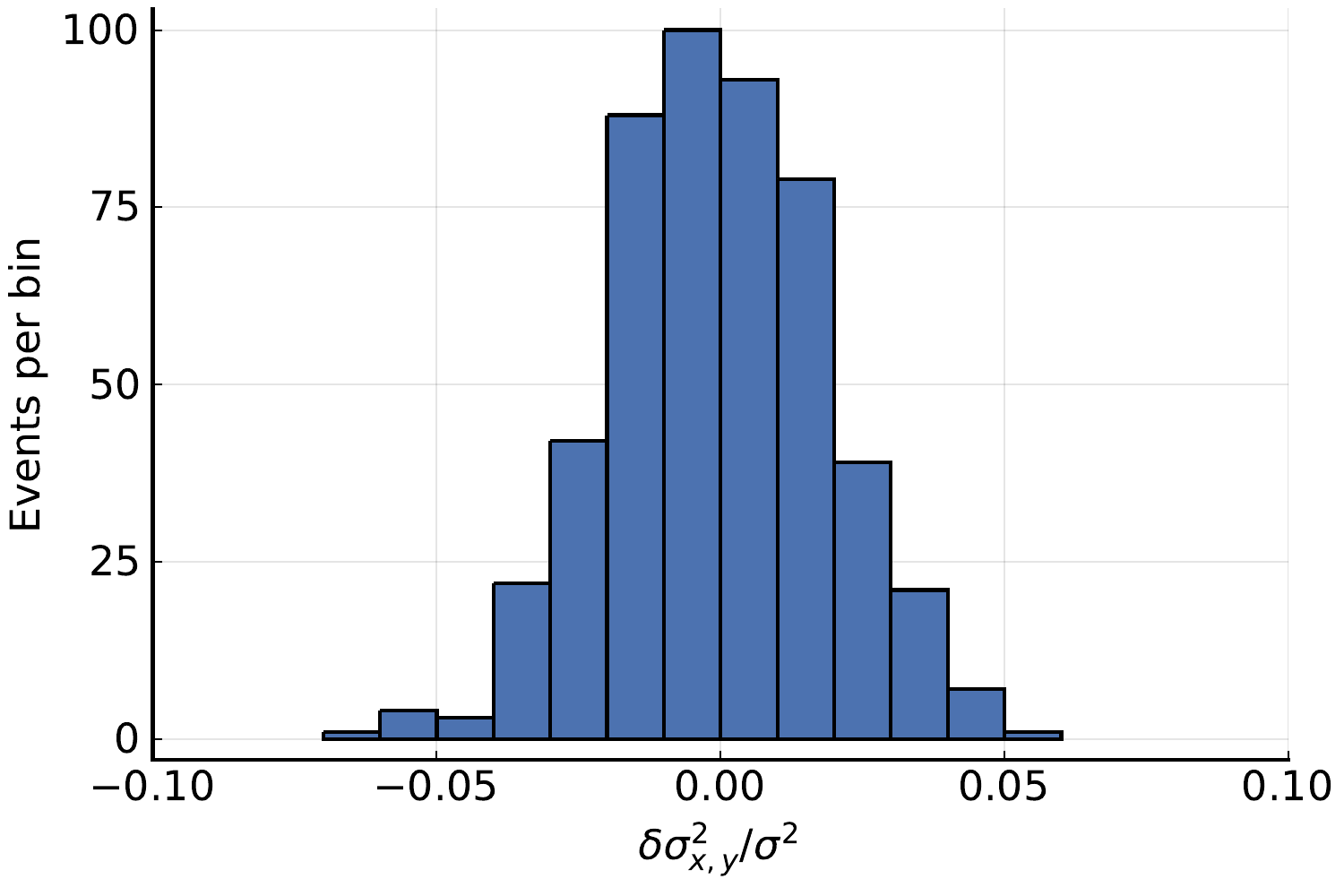}
	\caption{Histogram of beam size error as the strength of every quadrupole is varied randomly by $\sigma^{\Delta k/k}=0.1\%$. The histogram is obtained by five hundred single particle dynamics simulations with different seeds. In 95\% of the cases, the beam size error $\Delta \sigma_{x,y}^2/\sigma^2$ is within $\pm 3\%$.}
	\label{fig:all_quads_gaussian_offset}
\end{figure}

The focusing strength of each quadrupole can vary randomly as well. Assuming a
Gaussian variation with 0.1\% standard deviation ($\sigma^{\Delta
k/k}=10^{-3}$), we made 500 simulations with single particles to obtain the
histogram in Figure \ref{fig:all_quads_gaussian_offset}, which indicates that
the relative beam size variation is most likely within a few percent in this
scenario. In conclusion, the proposed BPM is tolerable to random quadrupole
strength drifts as long as they are within $\pm 0.1\%$, and one quadrupole
failure as long as the drift is less than 10\% level. 

\section {Profile reconstruction attempt using neural networks}\label{sec:neuralnetwork}

The neural network (NN) of choice was a dense \(\textbf{Y}\) to \(\textbf{X}\) fully connected
with a few layers.  A continuous funnel-shaped \(\textbf{Y}\) (8 dims) to
\(\textbf{X}\) ($20000=2\times 10000$ dims) with up to 10
fully connected layers NNs were tested. The size (both depth and width) did not
seem to affect the performance as much.

Keras~\cite{ref:keras} package was used to construct and train the
network. Standard procedures like L1,L2 normalization; early stopping; random
dropoff were implemented. The initial idea to use SSIM loss function came from various
modifications to NN setup. SSIM loss function, although costlier than MSE,
vastly outperformed MSE in terms of quality of the results.

Roughly $10^6$ (\(\textbf{X}\), \(\textbf{Y}\)) pairs were used in the training
stage. The training data consisted only from well-shaped 2-d gaussian beams (no
distortions). NNs perform acceptably when the training data and the testing data
are sampled from the same distribution (same data generating function). However,
when some distortion (bending the beams for example (as in Figure
\ref{fig:test_results})) is introduced (roughly in 10\% of the samples) in
either in the test or training samples, the results become unacceptable.
Simple-to-reconstruct shapes are misinterpreted as more complex shapes and
vice-versa. Absence of robustness despite the large training sample and
time-consuming training made us rethink the approach. The proposed genetic
algorithm requires no prior training (although fine-tuned beforehand); has
interpretable logic; is more performant.

Though convolution layers (or other popular choices like transformers) might
seem tempting to use, the problem at hand is reversed --- we need to go from low
dimensional data \(\textbf{Y}\) to high dimensional data \(\textbf{X}\).

A few additional details that are worth mentioning:

\begin{enumerate}
\item The order of the data point pairs in \(\textbf{X}\) is irrelevant. Any
permutation of the charged particle locations leads to the same signal. This
hints that usage of permutation invariant Graph Neural Networks or Set
Transformers might be possible as an extension for future studies. However, using
such state-of-the art models seems to be an overkill for such clear-stated
problem.

\item We expect physically relevant beam profile data \(\textbf{X}\) to be
continuous (without holes or tears) blobs with approximately Gaussian
distribution. To our knowledge, there does not seem to be an analytical,as well
as back propagation compatible, function to enforce this. However, the operators
(Section 3a) conserve such continuousness naturally.
\end{enumerate}
Another idea, was to use a somewhat middle ground approach --- Reinforcement
Learning (RL). RL  has not been attempted, but it could also be a natural step
toward future extension.

%\printbibliography

\end{document}